\documentclass[aps, prl,reprint, amsmath,amssymb, superscriptaddress,floatfix]{revtex4-1}
\usepackage{float}
\usepackage{graphicx}
\usepackage{xcolor,color,soul}
\usepackage{physics}
\definecolor{fb}{rgb}{0.45, 0.75, 0.26}
\definecolor{da}{rgb}{1, 0, 0}

\newcommand{\BSCCO}{Bi$_2$Sr$_2$CaCu$_2$O$_{8+\delta}$}
\usepackage{lineno}

\usepackage{graphicx}
\usepackage{dcolumn}
\usepackage{bm}
\usepackage{relsize}
\usepackage{xcolor,color}
\usepackage{ulem}
\usepackage{hyperref}
\hypersetup{colorlinks=true,urlcolor=blue,citecolor=blue}
\usepackage[symbol]{footmisc}

\begin{document}
\title{Direct evidence of light-induced phase-fluctuations in cuprates \\ via time-resolved ARPES}

\author{D. Armanno}
\altaffiliation{These authors equally contributed}
\affiliation{Advanced Laser Light Source, Institut National de la Recherche Scientifique, Varennes QC J3X 1P7 Canada}
\affiliation{Department of Physics, Center for the Physics of Materials, McGill University, 3600 rue Université, Montréal, Québec H3A 2T8, Canada}
\affiliation{Department of Chemistry, McGill University, 801 rue Sherbrooke Ouest, Montréal, Québec H3A 0B8, Canada}
\author{F. Goto}
\altaffiliation{These authors equally contributed}
\affiliation{Advanced Laser Light Source, Institut National de la Recherche Scientifique, Varennes QC J3X 1P7 Canada}
\author{J.-M. Parent}
\author{S. Lapointe}
\author{A. Longa}
\affiliation{Advanced Laser Light Source, Institut National de la Recherche Scientifique, Varennes QC J3X 1P7 Canada}
\author{R.\,D.\,Zhong}
\affiliation{Tsung-Dao Lee Institute $\&$ School of Physics and Astronomy, Shanghai Jiao Tong University, Shanghai 201210, China}
\affiliation{Condensed Matter Physics and Materials Science, Brookhaven National Laboratory, Upton, NY 11973, USA}
\author{J.\,Schneeloch}
\affiliation{Condensed Matter Physics and Materials Science, Brookhaven National Laboratory, Upton, NY 11973, USA}
\affiliation{Department of Physics $\&$ Astronomy, Stony Brook University, Stony Brook, NY 11795-3800, USA}
\author{G.\,D.\,Gu}
\affiliation{Condensed Matter Physics and Materials Science, Brookhaven National Laboratory, Upton, NY 11973, USA}
\author{G. Jargot}
\author{H. Ibrahim}
\author{F. L\'{e}gar\'{e}}
\affiliation{Advanced Laser Light Source, Institut National de la Recherche Scientifique, Varennes QC J3X 1P7 Canada}
\author{B.J. Siwick}
\affiliation{Department of Physics, Center for the Physics of Materials, McGill University, 3600 rue Université, Montréal, Québec H3A 2T8, Canada}
\affiliation{Department of Chemistry, McGill University, 801 rue Sherbrooke Ouest, Montréal, Québec H3A 0B8, Canada}
\author{N. Gauthier}
\affiliation{Advanced Laser Light Source, Institut National de la Recherche Scientifique, Varennes QC J3X 1P7 Canada}
\author{F. Boschini}
\email[]{fabio.boschini@inrs.ca}
\affiliation{Advanced Laser Light Source, Institut National de la Recherche Scientifique, Varennes QC J3X 1P7 Canada}
\affiliation{Quantum Matter Institute, University of British Columbia, Vancouver, BC V6T 1Z4, Canada}

\begin{abstract}
Phase fluctuations are widely accepted to play a primary role in the quench of the long-range superconducting order in cuprates. However, an experimental probe capable of unambiguously assessing their impact on the superconducting order parameter with momentum and time resolutions is still lacking. Here, we performed a high-resolution time- and angle-resolved photoemission study of optimally-doped Bi$_2$Sr$_2$CaCu$_2$O$_{8+\delta}$ and demonstrated a new experimental strategy to directly probe light-induced changes in the order parameter's phase with momentum resolution. To do this, we tracked the ultrafast response of a phase-sensitive hybridization gap that appears at the crossing between two bands with opposite superconducting gap signs. Supported by theoretical modeling, we established phase fluctuations as the dominant factor defining the non-thermal response of the unconventional superconducting phase in cuprates.
\end{abstract}

\maketitle
High-temperature copper-based superconductors, \textit{a.k.a.} cuprates, are a fascinating platform for exploring how strong electron correlations promote the emergence of intertwined quantum phases of matter \cite{keimer2017physics,davis2013concepts}. 
This complex interplay between different static and dynamic phases implies that these can be selectively favored or suppressed by tweaking the underlying energy landscape via external tuning knobs, such as magnetic field \cite{gerber2015three,chang2012direct}, twisted stacking \cite{cao2018unconventional,zhao2023time}, uniaxial strain \cite{bluschke2018stabilization,hicks2014strong} and light-matter interaction \cite{fava2024magnetic,fausti2011light,wandel2022enhanced} as non-exhaustive emblematic examples.  

In the specifics of light control of the superconducting (SC) phase, representative experimental achievements range from light-induced amplitude oscillations of the d-wave SC gap \cite{shimano2020higgs,chu2020phase}, as well as enhancement \cite{fausti2011light,fava2024magnetic} or quench \cite{boschini2018collapse,parham2017ultrafast,smallwood2012tracking,zhang2017photoinduced,wandel2022enhanced} of SC correlations.
In the last two scenarios, since the critical temperature $\text{T}_c$ of cuprates is dominated by the onset of the long-range phase coherence rather than the pairing strength \cite{emery1995importance,reber2012origin,kondo2015point,chen2022unconventional}, it is reasonable to expect that light excitation might have an (indirect) effect on the phase of the condensate \cite{boschini2018collapse,giannetti2016ultrafast,zhang2016stimulated,corson1999vanishing}. However, a momentum- and phase-resolved study that directly uncovers the emergence of phase fluctuations, as well as their impact on the superconducting condensate, is still missing.
 
Time- and angle-resolved photoemission spectroscopy (TR-ARPES), owing to its exquisite time, momentum and energy resolutions, grants direct access to how light excitation affects low-energy electrodynamics of quantum materials \cite{boschini2024time}. In particular, TR-ARPES enables the tracking of the ultrafast evolution of the SC gap \cite{smallwood2012tracking,smallwood2014time}, and current evidence is that light excitation with photon energies exceeding the SC gap amplitude transiently fills the SC gap \cite{boschini2018collapse,zhang2017photoinduced,zonno2021time,parham2017ultrafast}. This experimental observation has been discussed in terms of an ultrafast quench of the macroscopic phase coherence of the condensate (minimally affecting the pairing strength) via enhancement of phase fluctuations. 

However, since ARPES is not sensitive to the phase but only to the amplitude of the superconducting gap, one may wonder whether the transient filling of the SC gap is genuinely driven by phase fluctuations, or whether concomitant contributions acting on the electron self-energy (\textit{e.g.}, electron-electron and electron-phonon) may hinder this interpretation \cite{boschini2024time,zonno2021ubiquitous}. 
Recently, a new ARPES-based experimental strategy has been proposed to probe the phase of the SC gap \cite{gao2024arpes}. 
In particular, a (phase-sensitive) hybridization gap appears at the crossing of two bands only when they have SC gaps of opposite sign \cite{gao2024arpes,gao2020selective}. Moreover, since this hybridization gap is located well below the chemical potential (thus, more robust against thermal broadening and single-particle scattering contributions), the extension of this approach into the time domain promises a new avenue to directly and unambiguously unveil the role of phase fluctuations in the light-induced SC gap dynamics.

Here, we track the ultrafast evolution of the hybridization gap between the main and superstructure replica bands in \BSCCO (Bi2212) via TR-ARPES. We report a delayed ($\sim$1\,ps) quench of this phase-sensitive hybridization gap, with its subsequent recovery within a few ps. 
This new experimental evidence corroborates the interpretation of previous momentum-integrated and momentum-resolved time-resolved studies \cite{boschini2018collapse,zhang2017photoinduced,giannetti2016ultrafast,zonno2021time}, thus establishing that the contribution of phase fluctuations is essential for capturing the out-of-equilibrium response of superconducting cuprates.

\begin{figure}[t]
    \centering
    \hspace{-4mm}
    \includegraphics[width=0.43\textwidth]{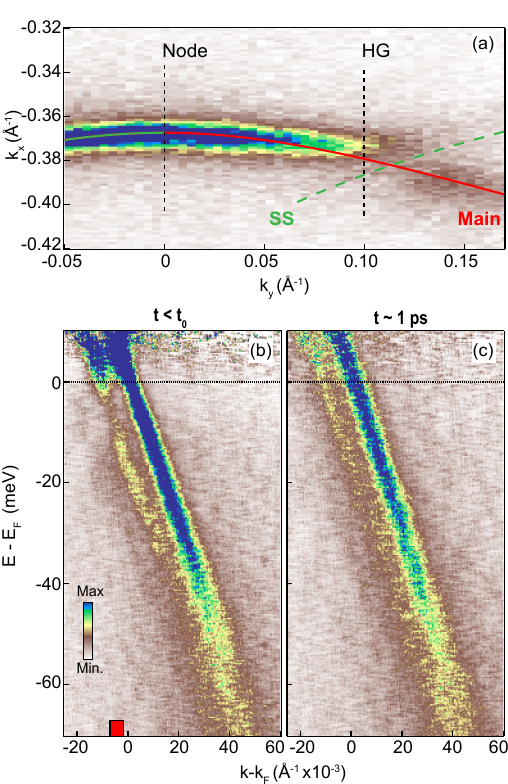}
    \caption{\textbf{Band mapping of the hybridization gap in Bi2212}
    (a) Isoenergy contour equilibrium mapping of Bi2212 ([-15,-5] meV integration window). The k$_{\text{x}}$ axis is aligned along the slit direction, and parallel to the $\Gamma - \text{X}$ direction. The solid and dashed lines indicate the tight binding constant energy contours of the main and superstructure replica (shifted by the superstructure vector $\textbf{Q}$=(0.254,0.254)\ $\text{\AA}^{-1}$) bands, respectively \cite{drozdov2018phase}. The green and red colors of the solid and dashed lines highlight the opposite sign of the SC gaps. 
    The vertical black dashed lines indicate the momentum cuts investigated in this work. 
    (b) and (c) ARPES maps of the hybridization gap (HG cut) before pump excitation (t$<$t$_0$) and at the maximum effect (t$\sim$1\,ps), respectively.
    ARPES maps have been deconvolved from both the energy resolution and impurity scattering contributions (details in the SM).
    }
    \label{Fig1}
\end{figure}

We performed TR-ARPES measurements on optimally doped Bi2212, with critical temperature $T_c \sim$91\,K, at the TR-ARPES endstation of the Advanced Laser Light Source (ALLS) user facility \cite{longa2024time}. The sample
was cleaved at a base temperature of 10\,K and pressure better than 8$\cdot$10$^{-11}$\,mbar, and aligned in-situ along the $\Gamma$-X direction. Electrons were photoemitted via s-polarized 6-eV probe pulses (electric field perpendicular to the analyzer's slit direction) and detected via an ASTRAIOS 190 (SPECS) hemispherical analyzer. Photoexcitation relied on mid-infrared  pump light (300\,meV photon energy, well below the charge transfer gap of Bi2212 \cite{cai2016visualizing}), at two different incident fluences, namely $\text{F}_1 \sim$ 25\,$\mu J/cm^2$ and $\text{F}_2 \sim$ 50\,$\mu J/cm^2$. The overall energy and temporal resolutions were 11\,meV and 350\,fs, respectively.

\begin{figure}[t]
    \centering
    \hspace{8mm}
    \includegraphics[width=0.47\textwidth]{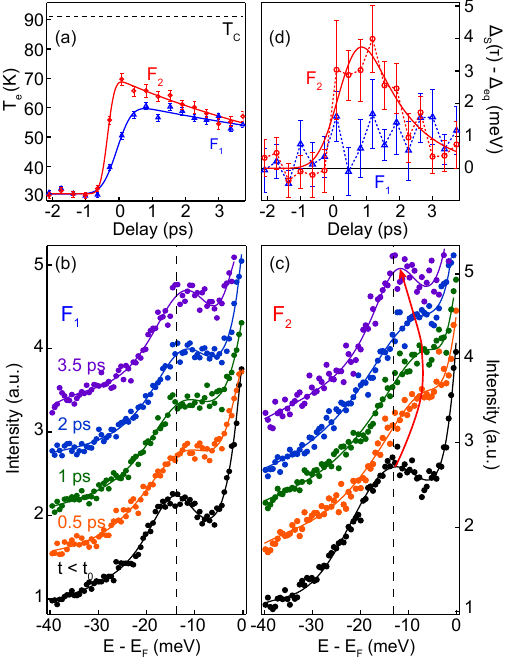}
    \caption{\textbf{Ultrafast dynamics of the hybridization gap.}
    (a) Transient evolution of the electronic temperature $T_e(t)$, extracted by fitting
    the momentum-integrated nodal EDC, for two fluences: $\text{F}_1$, blue and $\text{F}_2$, red. 
    (b) and (c) EDCs waterfall at the momentum indicated by the red square in Fig.\,\ref{Fig1}b at several pump-probe delays for $\text{F}_1$ and $\text{F}_2$, respectively. EDCs have been solely deconvolved from the energy resolution and divided by the Fermi-Dirac distribution evaluated using the electronic temperature of panel a. The EDCs were fit by a double-Lorentzian function (solid lines). The red solid arrow in c is a guide for the eye, and highlights the peak shift at $\text{F}_2$ fluence. 
    (d) Relative variation of the hybridization gap extracted from the fits of b and c (the equilibrium position of the peak is indicated by the black dashed lines in b and c). 
    Error bars represent the confidence interval in the fitting procedure corresponding to $\pm 2 \sigma$ ($\sigma$ is the standard deviation).
    }
    \label{Fig2}
\end{figure}

Figure\,\ref{Fig1}a displays the isoenergy contour equilibrium ARPES map at -10\,meV ($k_x$ aligned along the $\Gamma$-X direction), with the two momentum cuts investigated in this work: the nodal cut to extract the transient electronic temperature $T_e(t)$, and the hybridization gap (HG) cut. Note that, thanks to our hemispherical analyzer's deflector technology, the two momentum cuts were acquired in a single iterative delay scan without physically moving the sample, thus maintaining the same pump-probe interaction geometry. Therefore, the $T_e(t)$ estimated along the nodal direction (see Fig.\,\ref{Fig2}a) can also be used to reliably remove the thermal contribution from the photoemission intensity along the HG cut. 

The equilibrium ARPES map along the HG cut is presented in Fig.\,\ref{Fig1}b. In an effort to better visualize the underlying spectral features, we (i) deconvolved the energy distribution curves (EDCs) from the energy resolution \cite{yang2008emergence}, (ii) removed thermal broadening contributions to the ARPES intensity via Fermi-Dirac division, as well as (iii) deconvolved momentum distribution curves (MDCs) from an effective (energy and momentum independent) impurity scattering contribution (details in the SM).     
Since the anti-bonding main and the bonding superstructure replica (SS) bands have superconducting gaps of opposite sign in the momentum region where they cross (see Fig.\,\ref{Fig1}a), we observe the opening of a hybridization gap $\sim$10-15\,meV below the Fermi level (E$_{\text{F}}$), in perfect agreement with the recent ARPES work of Ref.\,\cite{gao2024arpes}. 

Interestingly, upon pump excitation (maximum effect at $\text{F}_2$ fluence, 1\,ps pump-probe delay, see Fig.\,\ref{Fig2}d), the hybridization gap vanishes, thus suggesting that light excitation affects the phase difference between the superconducting gaps of the main and SS bands. 
To directly connect the pump-induced quench of the hybridization gap to the transient enhancement of phase fluctuations, a detailed investigation of the temporal evolution of the hybridization gap spectral features is essential. The two fluence regimes employed in this work have been purposely selected to ensure that $T_e(t)$ is always well below the equilibrium $T_c$ (see Fig.\,\ref{Fig2}a), such that changes in the superconducting gap amplitude are negligible \cite{kondo2015point}. 

To track the transient evolution of the hybridization gap, we focus on the EDCs at the momentum indicated by the red square in Fig.\,\ref{Fig1}b. By simple visual inspection of the EDCs at negative delay (black markers in Fig.\,\ref{Fig2}b-c), we identify (moving down in binding energy from the E$_{\text{F}}$) the tail of a peak located above E$_{\text{F}}$ (see SM for further details), a dip at $\sim$-5\,meV, and finally a peak centered at $\sim$-15\,meV. This last peak captures the opening of the hybridization gap between the main and SS bands. 
Starting from the lowest fluence, $\text{F}_1$, Fig.\,\ref{Fig2}b reveals that pump excitation leads to a broadening of the spectral features but has a negligible impact on the peak position ($\sim$1\,meV shift, see Fig.\,\ref{Fig2}d). The spectral broadening is likely due to a transient increase in the single-particle scattering rate (that is linked to a higher electronic temperature), thus resulting in an apparent filling of the hybridization gap. 

The fluence $\text{F}_2$ instead shows a more peculiar behavior. Here, the hybridization gap peak undergoes a significant shift in energy upon light excitation of about $\sim$4\,meV. This transient shift is well captured by a delayed exponential function (red solid line in Fig.\,\ref{Fig2}d), peaked at $\sim$1\,ps and with a decay time of $\sim$1\,ps.  
Intriguingly, the timescales of the hybridization gap closure dynamics are reminiscent of what has been reported for the filling of the SC gap \cite{boschini2018collapse,zonno2021time,giannetti2016ultrafast}, \textit{i.e.} a delayed response with maximum effect reached after 0.5-1\,ps, and a decay time of the order of 1\,ps (see SM for an analysis of the SC gap evolution along the $\Gamma$-Y direction).
It is important to note that the temporal evolution of the hybridization gap (and pair-breaking scattering \cite{boschini2018collapse}) is markedly different from that of the electronic temperature, further demonstrating that the quench of the hybridization gap is non-thermal in nature.

In order to capture and model our experimental observation with the aim of verifying the claim that the transient quench of the hybridization gap is directly related to light-induced phase fluctuations, we expanded the phenomenological model proposed in Ref.\,\cite{gao2024arpes} by adding a random phase difference between the superconducting gaps of the main and SS bands.  
In particular, we consider the following 4x4 Hamiltonian capturing the interaction between two bands $\epsilon_{\alpha}$ and $\epsilon_\beta$ (tight binding model from Ref.\,\cite{drozdov2018phase}) 
\begin{equation}
    H = \begin{bmatrix}
        \epsilon_{\alpha} & V & \Delta_{\alpha}e^{i\phi_\alpha} &0\\[0cm]
        V & \epsilon_{\beta} & 0 & \Delta_{\beta}e^{i\phi_\beta}\\[0cm]
        \Delta_{\alpha}e^{-i\phi_\alpha} &0 & -\epsilon_{\alpha} & -V\\[0cm]
        0 & \Delta_{\beta}e^{-i\phi_\beta} & -V & -\epsilon_{\beta}
    \end{bmatrix},
\end{equation}
where $V$ is the coupling potential between the bands, and $\Delta_{\alpha}e^{i\phi_\alpha}$ and $\Delta_{\beta}e^{i\phi_\beta}$ are the two SC complex order parameters (in equilibrium conditions, $\phi_\beta-\phi_\alpha=\pi$).  

\begin{figure}[t]
    \centering
    \hspace{8mm}
    \includegraphics[width=0.45\textwidth]{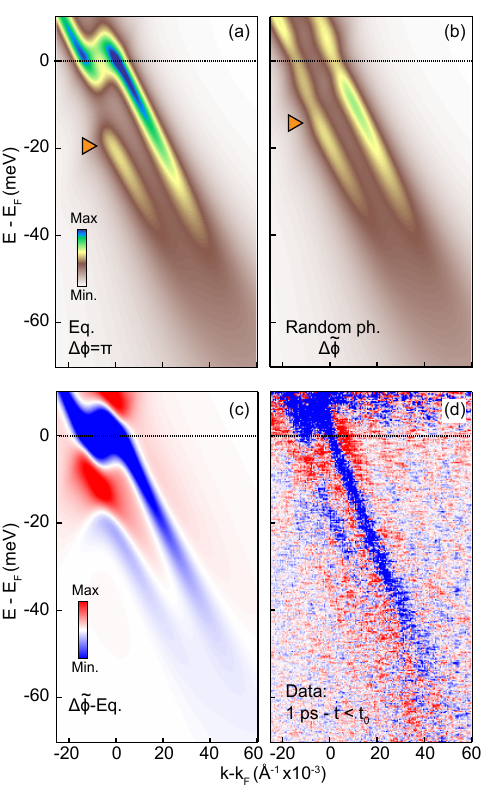}
    \caption{\textbf{Toy-model of the quench of the hybridization gap.}
    (a) and (b) Simulated ARPES maps with fixed $\pi$ and random phase difference, respectively. 
    (c) Differential ARPES map computed by subtracting a from b. (d) Experimental counterpart of panel c, computed using panels b and c of Fig.\,\ref{Fig1}. 
    }
    \label{Fig3}
\end{figure}

The results of these toy model simulations are displayed in Fig.\,\ref{Fig3}. The initial parameters of the simulation have been tuned to qualitatively reproduce the experimental ARPES map along the HG direction (Fig.\,\ref{Fig3}a with respect to Fig.\,\ref{Fig1}b). We then computed the ARPES map by averaging out the contribution of a random phase (phase difference ranging between $0$ and $2\pi$, see Fig.\,\ref{Fig3}b, details in SM). 
As a guide for the eye, the orange arrows in Fig.\,\ref{Fig3}a and b indicate the positions of the bottom branch of the hybridization gap and its shift upon pump excitation, respectively. Moreover, we compare the simulated and experimental differential ARPES maps in Fig.\,\ref{Fig3}c-d. Overall, this toy model captures remarkably well the main features of our experimental evidence. Indeed, (i) we reproduce a few meV shift of the $\sim$-15\,meV peak in the EDC crossing the hybridization gap, (ii) the hybridization gap is quenched only when the phase difference between the two bands departs from the initial value of $\pi$, as well as (iii) we report a qualitative agreement in the spectral weight increase/decrease of Fig.\,\ref{Fig3}c-d (note that since we removed the Fermi-Dirac contribution from the spectra, spectral weight changes are not caused by the broadening of the electronic distribution at high $T_e$). 

Based on the discussion above, we conclude that light excitation with photon
energy exceeding the SC gap amplitude quenches the hybridization gap via ultrafast enhancement of phase fluctuations, in good agreement with previous TR-ARPES works tracking the SC gap dynamics in the
near-nodal region \cite{boschini2018collapse,parham2017ultrafast,zhang2017photoinduced,smallwood2014time,zonno2021time}. In this regard, when comparing the hybridization gap dynamics of Fig.\,\ref{Fig2}d with that of the pair-breaking scattering rate
extracted from the SC gap filling analysis, we not only reproduce the $\sim0.5-1\,\text{ps}$ delayed response, but we also capture the $\sim1\,\text{ps}$ relaxation time. This implies a one-to-one correspondence between the pair-breaking scattering rate and hybridization gap dynamics, with the crucial difference that the quench of the hybridization gap is directly and solely related to the phase of the underlying SC condensate. We now have direct evidence that the light-induced enhancement of phase fluctuations is the main mechanism driving the quench of the SC condensate, and it has a non-thermal nature since $T_e$ is always well below $T_c$ and displays longer relaxation dynamics. 

The question remaining to be answered is how photoexcitation above the SC gap amplitude enhances phase fluctuations. The delayed response of the hybridization gap (and SC gap filling) points to the contribution of a non-thermal bosonic population \cite{coslovich2011evidence,giannetti2016ultrafast,boschini2018collapse,zhang2016stimulated}. To this end, an ultrafast inelastic electron diffraction study has reported a non-thermal low-energy optical phonon population during the first few ps after the photoexcitation \cite{konstantinova2018nonequilibrium}, and further time-resolved scattering experimental campaigns \cite{filippetto2022ultrafast,mitrano2020probing} may provide a definitive answer on how non-thermal bosons interact and quench the SC condensate. Furthermore, it is unclear whether light excitation disrupts the phase of the condensate within the CuO$_2$ plane (\textit{e.g.}, by introducing topological defects \cite{wandel2022enhanced,zong2019evidence}) or the phase between different CuO$_2$ (bi)layers \cite{hoegen2022amplification}, which in turn may lead to main and superstructure replica bands with an undefined phase difference. Nevertheless, the experimental approach proposed here promises to be applied to future TR-ARPES works with pump photon energies comparable to the SC gap amplitude to unveil whether the light-induced enhancement of SC correlations is driven by a transient locking of phase fluctuations \cite{giusti2019signatures,giannetti2016ultrafast,hoegen2022amplification}. Moreover, this new approach may 
clarify whether the ubiquitously reported $\sim$1\,ps relaxation dynamics is a measure of the intrinsic lifetime of the in-plane/inter-planes Cooper's pairs coherence, in both scenarios of light-induced quench or enhancement of SC correlations.		

\vspace{0.7 cm}
\begin{acknowledgments}
We thank the ALLS technical team for their support in the laboratory.
The work at ALLS was supported by the Canada Foundation for Innovation (CFI) -- Major Science Initiatives. We acknowledge support from 
the Natural Sciences and Engineering Research Council of Canada, the Canada Research Chairs Program (F.B.,B.J.S.), the CFI; the Fonds de recherche du Qu\'{e}bec --- Nature et Technologies, the Minist\`{e}re de l'\'{E}conomie, de l'Innovation et de l'\'{E}nergie --- Qu\'{e}bec, PRIMA Qu\'{e}bec, and the Gordon and Betty Moore Foundation’s EPiQS Initiative, grant GBMF12761 (F.B., F.L.). The work at BNL was supported by the US Department of Energy, office of Basic Energy Sciences, contract no. DOE-sc0012704.
\end{acknowledgments}

\clearpage
\onecolumngrid
\appendix

\renewcommand{\thefigure}{S\arabic{figure}}
\renewcommand{\theequation}{S\arabic{equation}}

\setcounter{figure}{0}
\setcounter{equation}{0}

\newpage

\section{Supplemental Material for \\ \textit{Direct evidence of light-induced phase-fluctuations in cuprates via time-resolved ARPES}}

\subsection{Data treatment and analysis}
The photoemission intensity can be written as 
\begin{equation}
I(\boldsymbol{k},\omega) = \left( |M|^2\cdot f(\omega) \cdot A(\boldsymbol{k},\omega)\right) \ast R(\omega) \ast K(k),
\end{equation}

where $|M|^{2}$ is the dipole matrix element, $f(\omega)$ is the electronic distribution (Fermi-Dirac distribution in equilibrium conditions), $A(\boldsymbol{k},\omega)$ is the one-electron spectral function and, finally, $R(\omega)$ and $K(k)$ are the experimental energy and momentum resolutions, respectively. 

\begin{figure*}[b]
    \centering
    \hspace{-0cm}
    \includegraphics[width=1\textwidth]{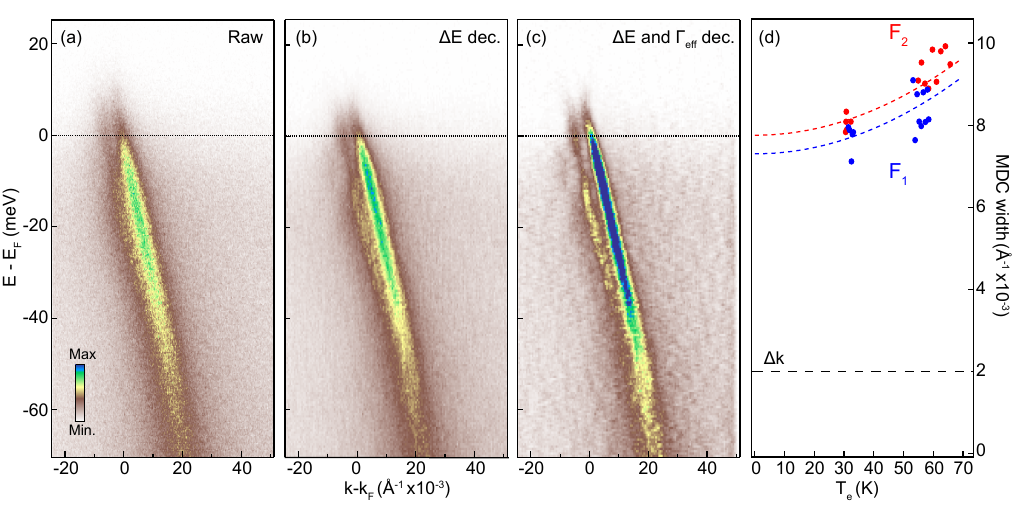}
    \caption{\textbf{Deconvolution effect on the ARPES maps.}
    (a) Raw ARPES map along the HG direction. (b) ARPES map in a, deconvolved from the energy resolution. (c) ARPES map in b deconvolved from the impurity scattering. (d) Nodal MDC width as a function of the electronic temperature $T_e(t)$, for $\text{F}_1$ (blue) and $\text{F}_2$ (red). Both curves are well captured by a Fermi-Liquid model (dashed lines). The black dashed line is the experimental (overestimated) momentum resolution in the experimental conditions of this work.
    }
    \label{FigS1}
\end{figure*}

The raw ARPES data along the HG cut of Fig.\,1 in the main text are shown in Fig.\,\ref{FigS1}a. Although the hybridization gap is already visible in the raw data, we deconvolved the experimental energy resolution $R(\omega)$, 11\,meV, from the photoemission intensity using the Lucy-Richardson algorithm \cite{yang2008emergence} to make the feature more evident (Fig.\,\ref{FigS1}b). 

In an effort to emphasize the hybridization gap even more (and without relying on methods such as 2D curvature, which only tracks the band dispersion but actually discards information about the underlying spectral function), we took the deconvolution procedure one step further. We deconvolved an additional broadening component given by impurity scattering. Note that since the momentum resolution (overestimated to be $\sim0.002\,\AA^{-1}$, dashed line in Fig.\,\ref{FigS1}d) is approximately 4 times smaller than the MDC width at the Fermi level (see Fig.\,\ref{FigS1}d), we neglect it.

The impurity scattering can be modeled as an $\omega$- and k-independent scattering rate $\Gamma_{eff}$ in the one-electron removal spectral function
\begin{equation}
A(k,\omega)=-\frac{1}{\pi}\frac{\Sigma^{''}(k)+\Gamma_{eff}}{(\omega-\epsilon_k-\Sigma^{'}(k))^2+(\Sigma^{''}(k)+\Gamma_{eff})^2}.
    \label{spectral}
\end{equation}

Under the assumption that the self-energy $\Sigma$ is k-independent, we realize that Eq.\,\ref{spectral} is the convolution of two Lorentzian functions of width $\Sigma^{''}$ and $\Gamma_{eff}$. 
We estimated $\Gamma_{eff}$ by tracking the temperature dependence of the nodal MDC at the Fermi level ($\pm 5 \text{meV}$, which is proportional to $\Sigma^{''}$), and extrapolating it to its zero temperature value. To do this, we used the transient electronic temperature extracted along the nodal direction, and assumed a Fermi-Liquid model, 
similar to what was done in Ref.\,\cite{zonno2021ubiquitous}.
The temperature dependence of $\Sigma^{''}$ is shown in Fig.\,\ref{FigS1}d, which is in good agreement with the parabolic fit of the Fermi Liquid model. We estimate $\Gamma_{eff}$= 7.5 and 7.8 $\times10^{-3}\AA ^{-1}$ for $F_1$ and $F_2$ fluence regimes, respectively. The final result of the Lorentzian deconvolution process (Lucy-Richardson algorithm) is shown in Fig.\,\ref{FigS1}c and Fig.\,1b-c of the main text. 
Note that only Fig.\,1b-c of the main text have been deconvoluted from the impurity scattering contribution, while all the other maps/EDCs of the main text have been solely deconvolved from the energy resolution.

\subsection{Ultrafast dynamics of the hybridization gap extended}

\begin{figure*}[t]
    \centering
    \includegraphics[width=0.7\textwidth]{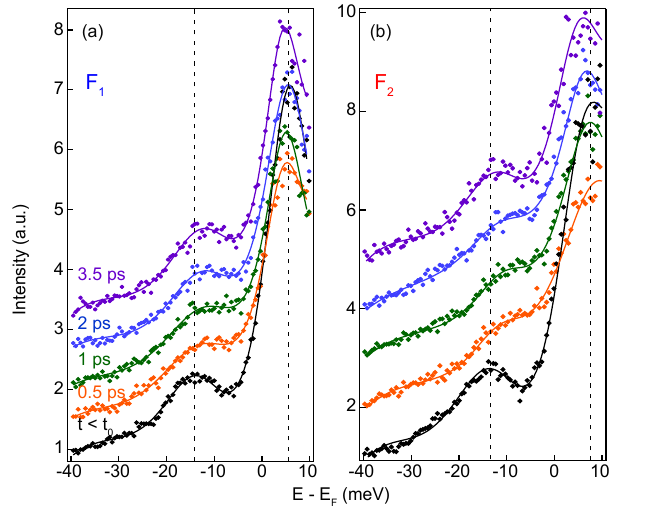}
    \caption{\textbf{Ultrafast dynamics of the hybridization gap - extended.}
    (a) and (b) EDCs waterfall at the momentum cut indicated by the red square in Fig.\,1b of the main text at selected pump-probe delays for $\text{F}_1$ and $\text{F}_2$, respectively. Coherently with the main text procedure, the EDCs have been deconvolved from the energy resolution, and divided by the Fermi-Dirac distribution evaluated using the relevant electronic temperature of Fig.\,2a. The EDCs were well fit by a double-Lorentzian fit (solid lines). 
    }
    \label{FigS2}
\end{figure*}

Here, we report the extended version (up to 10 meV above the Fermi level) of Fig.\,2 in the main text for the two fluences investigated. The EDCs cross the superstructure replica band for E \,$<0$ and the main band for E \,$>0$. 
Only the E \,$<0$ peak shifts as a function of pump-probe delay, and tracks the hybridization gap. Instead, the E \,$>0$ peak does not display any marked movements (Fig.\,\ref{FigS2}b).

\section{Superconducting gap sign simulation}

In order to reproduce the ARPES spectra presented in the main text, we simulate the one-electron removal spectral function
$A(k,\omega) = -\frac{1}{\pi}Im[G(k,\omega)]$
by using the Green's function
\begin{equation}
G(k,\omega) = (\omega-\Sigma(\omega)-H)^{-1},
\end{equation}

where $\Sigma(\omega)=i\Gamma+i\beta[\omega^2+(k_B\pi T)^2]$ is the electron self-energy expected from the Fermi-liquid model, and $H$ is the phenomenological Hamiltonian also reported in the main text:
\begin{equation}
    H = \begin{bmatrix}
        \epsilon_{\alpha} & V & \Delta_{\alpha}e^{i\phi_\alpha} &0\\[0cm]
        V & \epsilon_{\beta} & 0 & \Delta_{\beta}e^{i\phi_\beta}\\[0cm]
        \Delta_{\alpha}e^{-i\phi_\alpha} &0 & -\epsilon_{\alpha} & -V\\[0cm]
        0 & \Delta_{\beta}e^{-i\phi_\beta} & -V & -\epsilon_{\beta}.
    \end{bmatrix}
\end{equation}

Here, $\epsilon_a$ and $\epsilon_b$ are the bare dispersions of the main and superstructure bands, respectively, extracted from the tight-binding model of Ref\,\cite{drozdov2018phase}. The superstructure replica band is shifted by the superstructure vector $Q =(0.254,0.254)$. $\Delta_a$ and $\Delta_b$ represent the superconducting gap amplitude of the two bands and follow the typical d-wave momentum dependence. The initial parameters of the simulation are tuned to reproduce the momentum cut investigated in this work (Fig.\,1b) and they are reported in Supplementary Table\,\ref{tab1}.

\begin{table}[b]
    \centering
    \begin{tabular}{|| c | c | c | c | c | c | c | c ||}
    \hline
         $\mu$ (meV)&  $t$ & $t^{'}$ & $t^{''}$ & $t_{\bot}$ & $V$ & $\Delta_{0}$ & $Q$ ($\text{\AA}^{-1}$)\\ 
         \hline
         405 &  360 &  108 &  36 & 108 & 11 & 45 & 0.254
         \\
         \hline
    \end{tabular}
    \caption{Tight-binding and hybridization gap simulation parameters}
    \label{tab1}
\end{table}

\begin{figure*}
    \centering
    \hspace{-0mm}
    \includegraphics[width=0.8\textwidth]{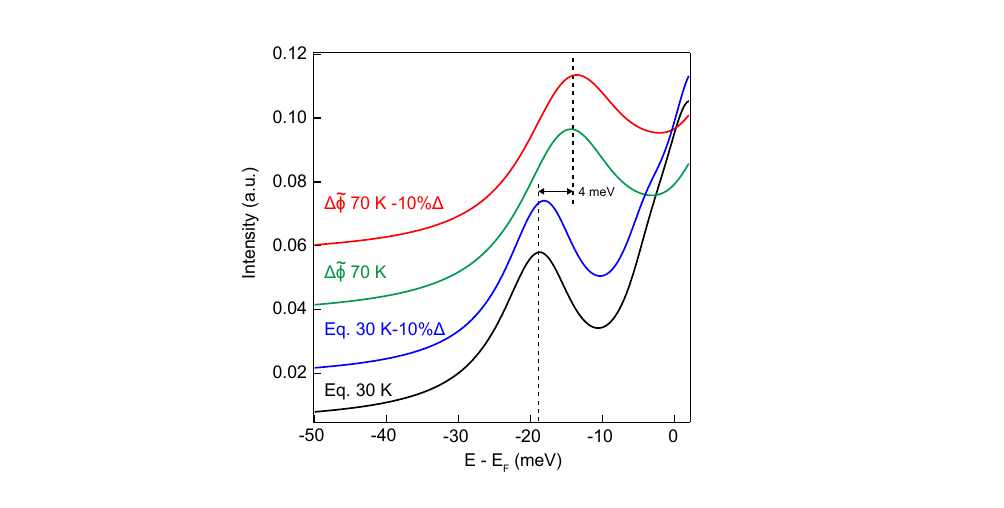}
    \caption{\textbf{Simulated equilibrium and random phase EDCs in different cases.}
    EDCs curves extracted from the integration range in Fig.\,1b of the main text for the following cases: (black) fixed phase difference at 30 K, (blue) fixed phase difference at 30 K with the SC gap reduced by 10$\%$, (green) random phase difference at 70 K, (red) random phase difference with a 10\% reduction of the SC gap amplitude at 70 K. 
    }
    \label{FigS3}
\end{figure*}

To further support our experimental findings, we report the simulated EDCs in Fig.\,\ref{FigS3}. The fixed phase case at $T=30 K$ represents the benchmark for this study, where $\phi_{\alpha}-\phi_{\beta}=\pi$ (black curve), and it displays a qualitatively good agreement with the negative delay EDC in the main text. 

The next step is to vary the phase difference of the two gaps to reproduce the ultrafast closing of the hybridization gap. 
We fix $\phi_{\alpha}$ and repeat the simulation varying $\phi_{\beta}$ between 0 and $2\pi$. By taking the average of the several phase differences (100 equally distributed values), we generate the green curve in Fig.\,\ref{FigS3} that qualitatively reproduces the energy shift observed in the data. 
For further accuracy, consistent with what has been reported for the SC gap closure at electronic temperature of about 70\,K \cite{kondo2015point}, we include a 10$\%$ reduction in the SC gap amplitude in the random phase-difference phase (red curve). However, this correction has a minimal effect on the actual position of the superstructure replica peak in energy. Therefore, our findings can be interpreted purely in terms of a light-induced enhancement of phase fluctuations of the SC gap. 
Finally, since we do not observe any light-induced change in the momentum splitting of the main and superstructure replica bands, we do not consider any transient change in the interaction strength $V$.

\subsection{Superconducting gap dynamics}

\begin{figure*}
    \centering
    \hspace{-0mm}
    \includegraphics[width=\textwidth]{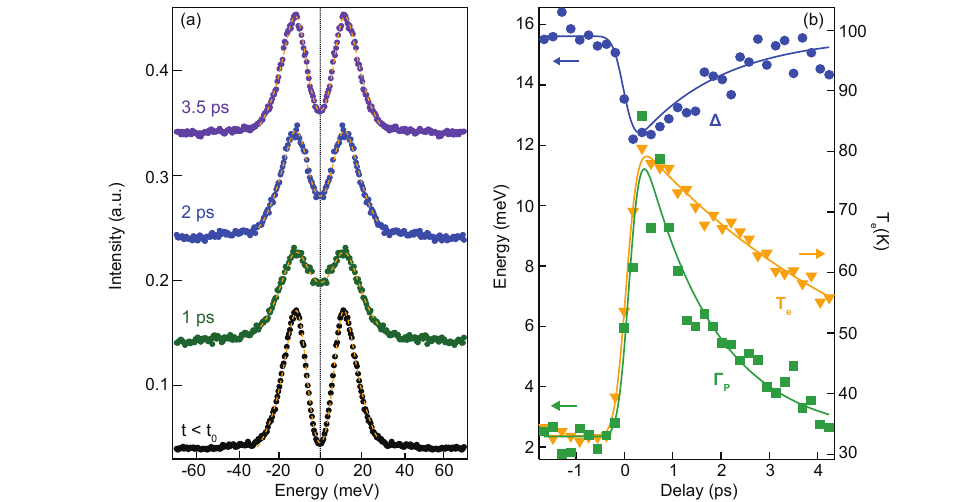}
    \caption{\textbf{Symmetrized EDCs at the Fermi momentum and ultrafast dynamics of the superconducting gap along $\Gamma$-Y.}
    (a) Symmetrized EDC (SEDC) curves extracted at the Fermi momentum $\text{k}_F$ for selected pump-probe delays. The orange dashed lines are the fit curves. (b)  Ultrafast dynamics of the superconducting gap and pair-breaking scattering rate (blue and green, respectively, extracted via Norman fit of the SEDCs in a), and transient electronic temperature (orange, extracted from the momentum-integrated nodal EDCs) as a function of the pump-probe delay. The solid lines are phenomenological single exponential decay fits.    
    }
    \label{FigS4}
\end{figure*}

Here we report 
the ultrafast dynamics of the superconducting gap along the $\Gamma -Y$ direction, \textit{i.e.} free from any contributions from the superstructure replica bands. We performed a pump-probe scan in this region with a slightly higher pump fluence ($\sim$75\,$\mu J/cm^2$), thus reaching a higher electronic temperature (80\,K instead of 70\,K for F$_2$ in the main text). 
In analogy to Ref.\,\cite{kondo2015point,boschini2018collapse}, we retrieve the transient evolution of the superconducting gap $\Delta$ and the pair-breaking scattering rate $\Gamma_\text{p}$ via Norman fit of the symmetrized EDCs (SEDCs) at the Fermi momentum $\text{k}_F$, as shown in Fig.\,\ref{FigS4}. The extracted dynamics are in perfect agreement with what has been observed in other works \cite{boschini2018collapse,zonno2021time}, confirming the non-thermal nature of $\Gamma_\text{p}$.
Instead, consistent with what has been reported by other equilibrium studies \cite{kondo2015point}, the SC gap amplitude reduces about $\sim$15/20$\%$ for the current fluence regime (higher than F$_2$ in the main text), and has a purely thermal origin \cite{zonno2021time}.

\newpage
\twocolumngrid
\providecommand{\noopsort}[1]{}\providecommand{\singleletter}[1]{#1}

\end{document}